\documentclass[preprint]{aastex}
\usepackage{graphicx}
\usepackage{txfonts}
\usepackage{url}

\newcommand{\rsun}{R$_{\odot}$}
\newcommand{\kms}{km~s$^{-1}$}

\newcommand{\dg}{^{\circ}}

\newcommand{\tii}{type~II}

\slugcomment{Submitted to ApJL on 4 July 2011. Accepted on August 9, 2011.}

\shorttitle{Super- and Sub-critical Regions in CME-driven Shocks}
\shortauthors{Bemporad \& Mancuso}

\begin{document}

\title{Identification of Super- and Sub-critical Regions\\
	  in Shocks driven by Coronal Mass Ejections}

\author{A. Bemporad \& S. Mancuso}
\affil{Istituto Nazionale di Astrofisica (INAF), Osservatorio Astronomico di Torino, \\ Strada Osservatorio 20, 10025 Pino Torinese, Torino, Italy} 
\email{bemporad@oato.inaf.it}

\begin{abstract}
In this work, we focus on the analysis of a CME-driven shock observed by {\it SOHO}/LASCO. We show that white--light coronagraphic images can be employed to estimate the compression ratio $X=\rho_d/\rho_u$ all along the front of CME--driven shocks. $X$ increases from the shock flanks (where $X \simeq 1.2$) to the shock center (where $X \simeq 3.0$ is maximum). From the estimated $X$ values, we infer the Alfv\'en Mach number for the general case of an oblique shock. It turns out that only a small region around the shock center is supercritical at earlier times, while higher up in the corona the whole shock becomes subcritical. This suggests that CME-driven shocks could be efficient particle accelerators at the initiation phases of the event, while at later times they progressively loose energy, also losing their capability to accelerate high energy particles. This result has important implications on the localization of particle acceleration sites and in the context of predictive space weather studies.
\end{abstract}

\keywords{Sun: coronal mass ejections (CMEs); shock waves; }

\section{Introduction}
Solar energetic particle (SEP) events represent one of the most severe hazards in the near--Earth space environment. The particle population of SEP events at Earth critically depends on the location of the particle acceleration, which may occur in both the solar corona and interplanetary space. While the association between shocks driven by Coronal Mass Ejections (CMEs) and gradual SEP events is clear, the acceleration mechanisms underlying this association is less understood (Zank et al. 2000). In general, shocks take the flow energy upstream and convert it to thermal energy downstream and accelerated particles. 
Only supercritical shocks, i.e., shocks with Alf\'enic Mach number $M_{\rm A}$ (the ratio of the upstream flow speed along the shock normal to the upstream Alfv\'en speed) $> M_{\rm A}^\star$ (the first critical Mach number) are thought to be able to reflect ions and thus produce Sjpg close to the Sun ($< 3$ \rsun). In fact, above $M_{{\rm A}}^\star$, resistivity alone is unable to provide all the dissipation needed for the required Rankine-Hugoniot shock jump  (e.g., Edmiston \& Kennel 1984). 
Critical Mach numbers range between 1 and 2 for quasi-parallel shocks and between 2.3 and 2.7 for quasi-perpendicular shocks for the coronal plasma, depending on both $\theta_{Bn}$ and the plasma $\beta$ of the ambient medium through which the shock propagates.
As supercriticality is surpassed, ion reflection is required to provide the required downstream heating (e.g., Burgess 2006).
Supercritical shocks are also thought to be required for accelerating electrons in the energy range relevant to the production of \tii\ bursts, which are the radio signature of coronal shock waves (Gopalswamy et al. 2010).  
Large  increases in electron fluxes in the energy range 2 -- 20 keV, associated with interplanetary shocks  with high Mach number have been actually reported (e.g., Tsurutani \& Lin 1985) and the fact that
type II radio bursts also show considerable variability may be due to the associated shock changing Mach number as it propagates through an inhomogeneous solar wind (e.g., Knock et al. 2003). 
Thus far, however, the clear identification of the shock--associated particle acceleration region  in the corona and its evolution with time is still missing, although some progress has been made in the identification of candidate shock signatures both in whit-- light images (e.g., Ontiveros \& Vourlidas 2009; Liu et al. 2009) and in ultraviolet spectra (e.g., Raymond et al. 2000; Ciaravella et al. 2005; Mancuso \& Avetta 2008; Bemporad \& Mancuso 2010; Mancuso 2011).

In this Letter, we analyze white-light observations of a fast CME-driven shock observed by the LASCO instrument on board {\it SOHO}. From the observed compression ratio along the shock front, we estimate $M_{\rm A}$ as a function of latitude and height in the corona, showing that the shock actually becomes supercritical, and thus more efficient in accelerating both electrons and ions to high energies and, correspondingly, in generating high fluxes of energetic particles, only at selected angles around the shock center. This region is moreover seen to disappear as the shock propagates in the coronal environment, thus implying a lower ability in accelerating Sjpg at progressive heights.
\begin{figure}[t]
\begin{center}
  \includegraphics[height=.40\textheight]{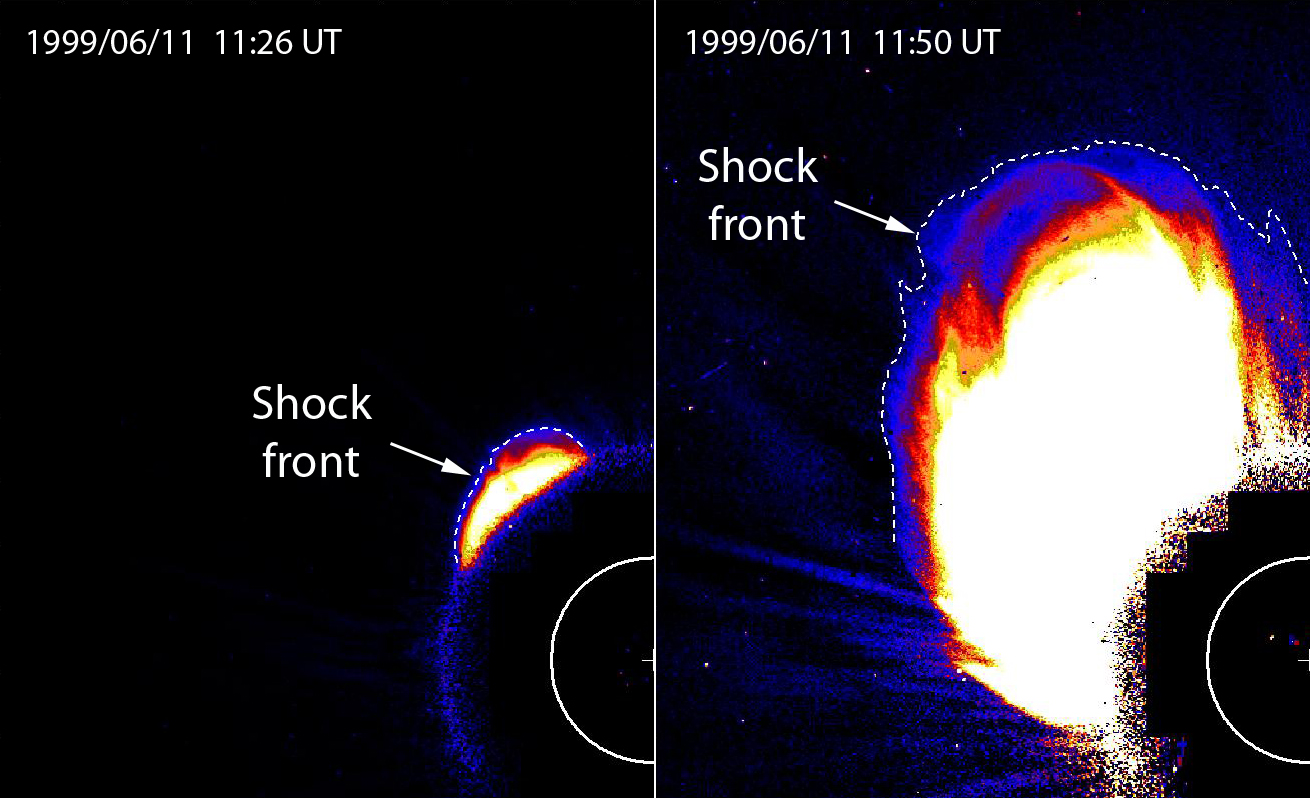}
  \caption{Base difference LASCO/C2 images showing the location of the CME-driven shock front ({\it dashed line}s) at 11:26 UT ({\it left}) and 11:50 UT ({\it right}).}\label{fig1}
\end{center}
\end{figure}
\begin{figure*}[th]
\begin{center}
  \includegraphics[height=.45\textheight]{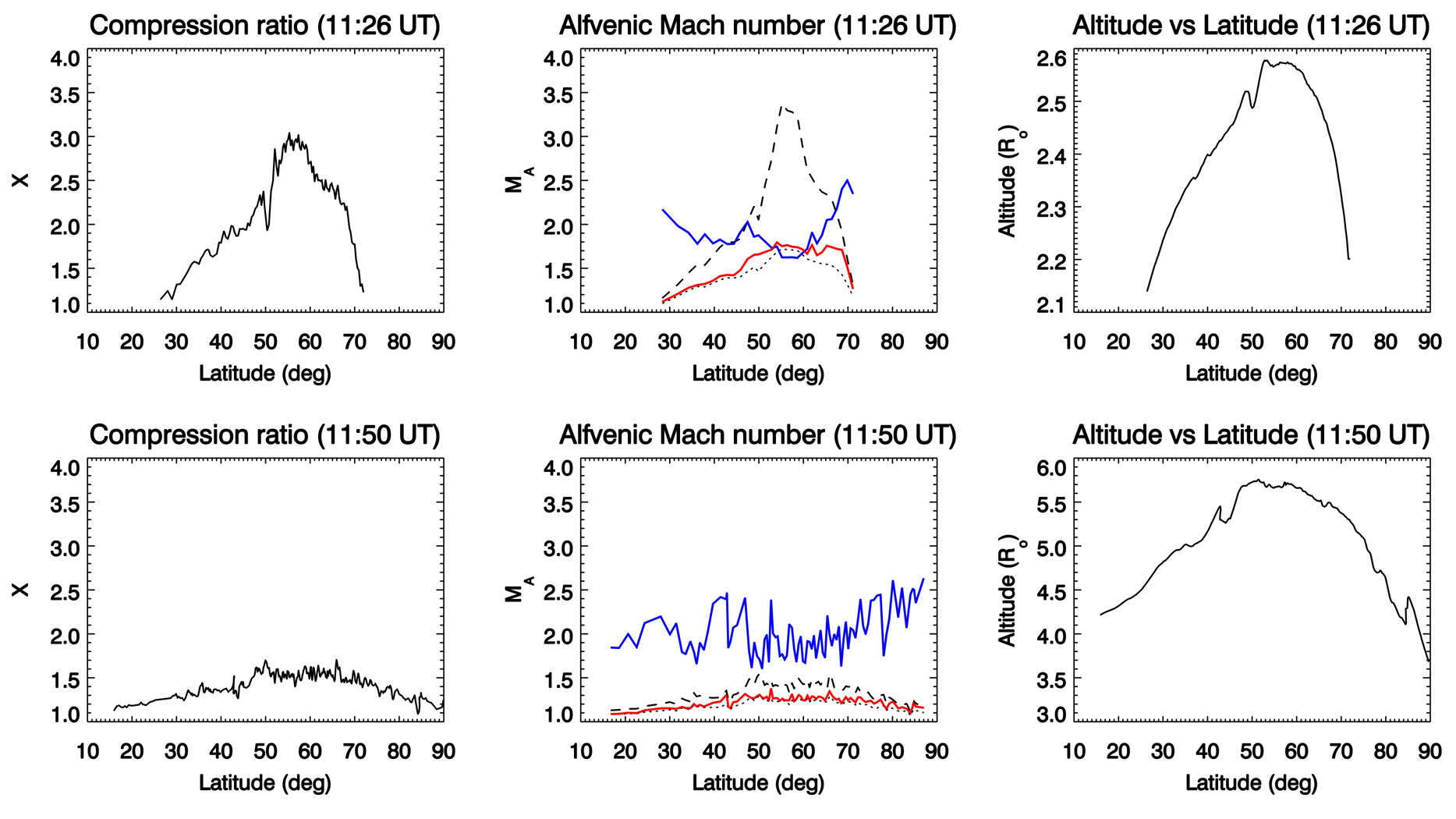}
  \caption{{\it Left}: Compression ratios $X\equiv\rho_{\rm d}/\rho_{\rm u}$ as measured at the points illustrated by the dashed lines in Figure 1 along the shock fronts at 11:26 UT ({\it top}) and 11:50 UT ({\it bottom}). {\it Middle}: Alfv\'en Mach numbers $M_{\rm A}$ for perpendicular ({\it dashed line}) and parallel ({\it dotted line}) and for angles measured along the actual shock front ({\it solid red lines}) at 11:26 UT ({\it top}) and 11:50 UT ({\it bottom}); the latter curves are compared with the corresponding critical Mach numbers ({\it solid blue lines}). {\it Right}: Altitudes of points located at different latitudes along the shock front in the previous plots.}\label{fig2}
\end{center}
\end{figure*}

\section{Observations and data analysis}

Figure 1 shows {\it SOHO}/LASCO C2 base difference images of the 1999 June 11 event. This CME was first detected at 11:26 UT in the LASCO C2 field of view below a height of 2.6 \rsun, travelling in the Northeast direction at a speed of 1569 \kms, as provided by the online LASCO CME catalog (Gopalswamy et al. 2009). According to  {\it GOES} data, the soft-X ray emission of the associated C8.8 flare started at 11:07 UT and reached a maximum at 11:57 UT. The flare was emitted from active region NOAA 8585, which was located at a latitude of 38$\dg$ above the equator and just a few degrees behind the limb. Radio images acquired $\sim 6$ minutes after the flare onset by the Nan\c cay Radioheliograph (NRH, Radioheliograph Group 1989) show the formation of an arch-shaped radio emission above the limb at the latitude of the CME source region. The 164 MHz emission detected by NRH, which is simultaneous to the \tii\ radio signature observed in ground-based metric dynamic spectra, is strongly suggestive of the plane-of-the-sky section of a CME-driven coronal bow-shock surface. 

The CME front was observed in only two LASCO C2 frames acquired at 11:26 and 11:50 UT.  The derivation of the electron density from the polarized brightness ($pB$), which is photospheric white--light scattered off coronal electrons, is a classical problem (Van de Hulst 1950). By assuming spherical symmetry, we fitted two-component radial power laws to the observed pre-CME $pB$ radial profiles, from which we obtained a set of coronal electron density profiles all over the region of shock propagation. In order to determine the location of the shock front, identified as a weak white--light intensity increase located above the expanding CME front (Figure 1, {\it dashed lines}), we subtracted the average pre-CME intensity from each frame (see also Ontiveros \& Vourlidas 2009 for a discussion on white--light shock identification). For each pixel along the front, we extracted the unpolarized white--light brightness of the shocked plasma, $b_{\rm post}$, while the last image acquired before the CME arrival provided the pre-shock brightness, $b_{\rm pre}$, at the same locations in the corona. In each pixel, the brightness $b(\zeta)$ observed at a projected altitude $\zeta$ is given by a line-of-sight (LOS) integration of the inferred electron density profile, $n_e$, multiplied by a geometrical function. The integration along the LOS is then divided into two integrals: one performed over the unshocked corona (with density $n_e$) and the other over a length $L$ across the shocked plasma with density $X\cdot n_e$. Here, $X\equiv\rho_{\rm d}/\rho_{\rm u}$ is the unknown shock compression ratio and $\rho_{\rm d}$ ($\rho_{\rm u}$) is the downstream (upstream) plasma density. In this work $L$ has been estimated as in Bemporad \& Mancuso (2010), i.e. by assuming that the shock surface has the 3-D shape of an hemispherical shell with thickness equal to that observed on the plane of the sky, corrected for the shock motion during the LASCO/C2 exposure time. Resulting values are $L=0.28$ R$_\odot$ and $L=0.61$ R$_\odot$ for the shock observed at 11:26 and 11:50 UT, respectively. Given $L$, by adopting the radial density profile derived from the analysis of the $pB$, the shock compression ratio $X$ can be inferred from the observed $b_{\rm pre}$ and $b_{\rm post}$.

The so-obtained compression ratios $X$ along the shock fronts at 11:26 and 11:50 UT, are shown in Figure 2 ({\it left panels}). The $X$ values vary between 1.2 and 3.0, a range that is lower than the upper limit compression of 4 given by $X = (\gamma+1)/ (\gamma-1)$ for a mono-atomic gas ($\gamma = 5/3$). An interesting new result is given by the latitudinal dependence: at both times $X$ maximizes at the center of the shock surface, progressively decreasing towards the flanks of the shock. Notice that values plotted in this Figure may also depend on unknown latitudinal changes in the $L$ values, which have been assumed constant at all latitudes. A second new result is the time evolution: as the shock expands, the latitudinal dependence is preserved, but the $X$ values decrease all along the shock surface. This means that the shock is losing its energy with propagation and that, at least for this event, it is stronger at the center of the front, as one would expect because this is the shock part moving faster in the corona, hence with larger $M_A$. The latter conclusion is demonstrated in the next section.

The relationship between the compression ratio $X$ and the Alfv\'enic Mach number $M_{\rm A}$ depends on the plasma-to-magnetic pressure ratio $\beta$ and the angle $\theta_{Bn}$ between the shock normal and the upstream magnetic field. In the solar corona, the plasma is mainly controlled by the magnetic field ($\beta\ll 1$). In the limiting case of $\beta\rightarrow 0$, the Alfv\'enic Mach number is given by $M_{{\rm A}\perp}=\sqrt{X(X+5)/[2(4-X)]}$ for a perpendicular shock and $M_{{\rm A}\parallel}=\sqrt{X}$ for a parallel shock (e.g., Vr{\v s}nak et al. 2002). The angles $\theta_{Bn}$ have been estimated pixel by pixel along the shock surfaces from the LASCO images by assuming a radial magnetic field above $\sim 2$ \rsun. By taking, as a first order approximation for the more general case of an oblique shock, $M_{{\rm A}\angle}=\sqrt{(M_{{\rm A}\perp}\sin\theta_{Bn})^2+(M_{{\rm A}\parallel}\cos\theta_{Bn})^2}$, implying $M_{{\rm A}\perp} \geq M_{{\rm A}\angle} \geq M_{{\rm A}\parallel}$, we finally obtained the values plotted in Figure 2 ({\it  middle panels}). $M_{{\rm A}\angle}$ is seen to maximizes at the shock center at both times, reaching a maximum value of $M_{{\rm A}\angle}\simeq 1.8$ at $\sim 2.6$ \rsun. Critical values for the Alfv\'enic Mach number $M_{{\rm A}}^\star$ for collisionless non-relativistic shocks are provided by Edminston \& Kennel (1984), recently reviewed by Treumann (2009). The above authors provide a theoretical $M_{{\rm A}}^\star$ curve for a fast-mode shock as a function of the angle $\theta_{Bn}$ for $\beta = 0$. From the $\theta_{Bn}$ values derived along the shock front, we estimated $M_{{\rm A}}^\star$ at different latitudes. The resulting $M_{{\rm A}}^\star$ variations along the front are seen to be opposite to those of $M_{{\rm A}\angle}$. That is, $M_{{\rm A}}^\star$ is seen to minimize at the shock center and maximize at the flanks. Comparison between the $M_{{\rm A}\angle}$ and $M_{{\rm A}}^\star$ curves in Figure 2 shows that, at earlier times and lower altitudes, $M_{{\rm A}\angle} > M_{{\rm A}}^\star$ at the center of the shock and $M_{{\rm A}\angle} < M_{{\rm A}}^\star$ at the shock flanks, while, at later times and higher altitudes, $M_{{\rm A}\angle} < M_{{\rm A}}^\star$ at all latitudes. Hence, at earlier times the shock center is super-critical, making it a probable source for SEP acceleration.
\begin{figure}[th]
\begin{center}
  \includegraphics[height=.40\textheight]{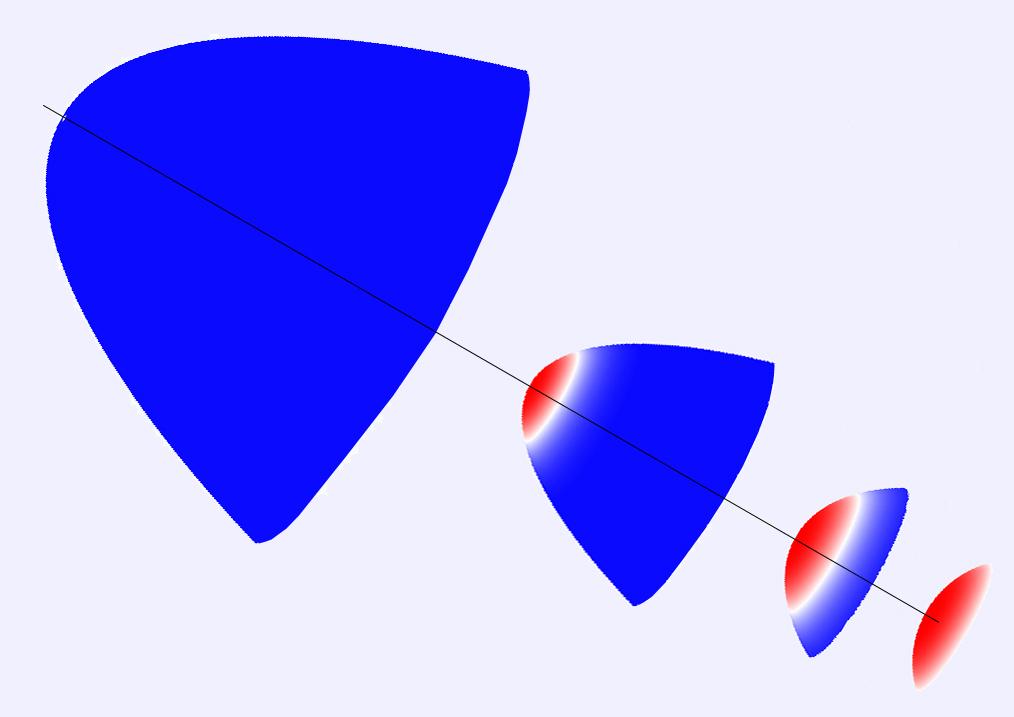}
  \caption{Cartoon showing the possible evolution of supercritical ({\it red}) and subcritical ({\it blue}) regions over the shock surface.}\label{fig3}
\end{center}
\end{figure}
\section{Discussion and conclusions}

In this work, we have shown for the first time that coronagraphic white--light images can be employed to estimate the shock compression ratio $X$ not only at a single latitude and altitude (as in previous works), but all over the observed shock front. The compression ratios computed for an event observed by {\it SOHO}/LASCO-C2 have been used to estimate the Alfv\'enic Mach numbers for perpendicular, parallel and oblique shocks. The results show that at earlier times the CME-driven shock is supercritical at the top of the shock and subcritical at the flanks, with the supercritical region disappearing with time. The above results suggest the overall evolution qualitatively shown in Figure 3, with the CME-driven shock primarily super-critical in its earlier phase (and hence an efficient particle accelerator). In the following minutes, as the shock progressively lose its energy, the supercritical region becomes narrower and narrower, with the particle acceleration site progressively more localized in space. At the end of the process, the whole shock becomes subcritical, so that no more efficient high energy particle acceleration is possible. While the present result for a single event supports SEP generation at the CME/shock front in the higher corona, type II emission at the CME flanks is still expected in the lower corona due to interaction with nearby streamers (e.g., Mancuso \& Raymond 2004), so that a more extended study will be required in order to draw a definite conclusion. 
These results have very important implications on the localization of particle acceleration sites during CMEs and on the temporal evolution of SEP fluxes in CME--driven shocks.

\acknowledgments
A.B. acknowledges support from ASI/INAF I/023/09/0 contract. SOHO is a project of international cooperation between ESA and NASA.


\begin{thebibliography}{}
\bibitem[]{}Bemporad, A., \& Mancuso, S. 2010, \apj, 720, 130
\bibitem[]{}Brueckner, G. E, et al. 1995, \solphys , 162, 357
\bibitem[]{}Burgess, D. 2006, \apj, 653, 316
\bibitem[]{}Ciaravella, A., Raymond, J.~C., Kahler, S.~W., Vourlidas, A., \& Li, J.\ 2005, \apj, 621, 1121 
\bibitem[]{}Edmiston, J. P., \& Kennel, C. F. 1984, J. Plasma Phys., 32, 429
\bibitem[]{}Gopalswamy, N., Yashiro, S., Michalek, G., et al. R. 2009, Earth Moon and Planets, 104, 295 
\bibitem[]{}Knock, S. A., Cairns, I. H., Robinson, P. A., \& Kuncic, Z. 2003, \jgr, 108, 6
\bibitem[]{}Liu, Y., Luhmann, J. G., Bale, S. D., \& Lin, R. P. 2009, \apj, 691, L151
\bibitem[]{}Mancuso, S., 2011, \solphys, in press  
\bibitem[]{}Mancuso, S., Avetta, D. 2008, \apj, 677, 683      
\bibitem[]{}Mancuso, S. \& Raymond, J. C. 2004, \aap, 413, 363
\bibitem[]{}Ontiveros, V. \& Vourlidas, A. 2009, \apj, 693, 267            
\bibitem[]{}Raymond, J. C., et al. 2000, \grl, 27, 1439
\bibitem[]{}Radioheliograph Group 1989, \solphys, 120, 193 
\bibitem[]{}Treumann, R. A. 2009, Astron. Astrophys. Rev., 17, 409
\bibitem[]{}Tsurutani, B. T., \& Lin, R. P. 1985, \jgr, 90, 1
\bibitem[]{}van de Hulst, H. C. 1950, Bull. Astron. Inst. Netherlands, 11, 135
\bibitem[]{}Vourlidas, A., Wu, S. T., Wang, A. H., Subramanian, P., \& Howard, R. A., 2003, \apj, 598, 1392     
\bibitem[]{}Vr{\v s}nak, B., Magdaleni{\'c}, J., Aurass, H., \& Mann, G.\ 2002, \aap, 396, 673 
\bibitem[]{}Zank, G., Rice, W. K. M., \& Wu, C. C., 2000, \jgr, 105, 25079                     
\end{thebibliography}
\end{document}